# Multidimensional Double Refraction Microscopy

J. König[1,2]* and C. Cierpka[1,2]

[1]Institute of Micro- and Nanotechnologies, Technische Universität Ilmenau, Germany

[2]Institute of Thermodynamics and Fluid Mechanics, Technische Universität Ilmenau, Germany

*e-mail: Joerg.Koenig@tu-ilmenau.de

**Abstract**

Multidimensional optical microscopy – extracting 3D location, orientation, and spectral information from nanoscale emitters – is key to modern nanoscience. However, this multidimensional capability requires complex instrumental setups, demanding exceptional stability and specialized expertise in optics. Here, we overcome these barriers by introducing Double Refraction Microscopy (DRM), a passive imaging method providing a straightforward pathway to multidimensional super-resolution imaging. Based on bifocal imaging via a birefringent microscope slide, DRM splits nanoprobe light into a characteristic double image encoding 3D location, orientation, and spectral signature, while requiring zero hardware modifications to standard wide-field microscopes. To showcase its robustness, we implement DRM across three distinct slides of varying materials and thicknesses, calibrating 3D localization with 450-nm nanoparticles and demonstrating high orientation sensitivity using polarization-controlled emitters. Furthermore, we confirm spectral readout using a pinhole array acting as a grid of point emitters back-illuminated at various wavelengths to generate distinct spectral signatures. Finally, we showcase practical utility by performing a two-colour 3D localization measurement within a 10-µm gap, utilizing the single optical pathway of a standard epi-fluorescence microscope. Ultimately, DRM provides the missing optical hardware simplicity to complement established computational tools, promising to transfer advanced super-resolution capabilities from specialized facilities to every laboratory bench.

## INTRODUCTION

More than three decades of advancements in wide-field optical microscopy, particularly in Single-Particle Tracking (SPT)[1–3] and Super-Resolution Microscopy (SRM)[1,4,5] have revolutionized our ability to observe nanoscale structural architectures and dynamic processes. Originally enabled by switching fluorophores between emissive and dark states[6,7] in stochastic approaches like PALM[8] or STORM[9], these localization frameworks have found widespread adoption across diverse disciplines. These range from resolving nanoscale architectures in cell biology[10,11] to probing complex transport phenomena in electrochemistry[4,12], soft matter physics[5], and microfluidics[1,2]. However, comprehensively characterizing these complex synthetic and biological systems increasingly demands multidimensional imaging capabilities that can simultaneously capture the three-dimensional (3D) position, the molecular orientation, and the spectral signature of individual emitters[13,14]. Scaling optical detection to these multiple dimensions typically introduces severe technical trade-offs. Axially resolved 3D localization routinely requires acquiring information from different depths using multi-focal plane microscopy[15–17] or breaking the axial symmetry of the point spread function (PSF) via astigmatism[10,18] or advanced phase masks in the back focal plane, such as the double-helix[19] or Tetrapod PSFs[20]. To additionally extract the azimuthal ($\varphi$) and polar ($\theta$) orientation of dipole emitters[21–24], state-of-the-art modalities rely on polarization-based splitting into orthogonal channels[25] or complex PSF-engineering frameworks like the bisected pupil[26], tri-spot PSF[27], or CHIDO[28] to approach fundamental precision limits[29]. When spectral discrimination is further introduced – either through multi-channel ratiometric splitting[30,31] or dispersive single-path phase masks[32,33] – the instrumental complexity scales drastically. These traditional multi-path architectures severely suffer from compromised photon budgets due to channel splitting[34], reduced fields of view (FOVs)[30,32], and the necessity for rigorous calibration and image registration[1,17,34,35]. Consequently, despite the solid foundation of robust optical probes[36] and sophisticated computational frameworks[37–39], the underlying hardware complexity remains a significant barrier. The reliance on delicate optical alignments, customized spatial light modulators (SLMs), and multi-camera frameworks currently restricts truly multidimensional imaging to specialized optical laboratories and core facilities[14,34,40]. There is a clear need for straightforward, robust, and single-path optical modalities that can democratize multidimensional particle tracking and imaging without compromising photon efficiency or alignment simplicity.

Here, we introduce a single-path optical imaging method that enables the simultaneous measurement of the 3D position, dipole orientation, and spectral signature of individual emitters. Remarkably, our approach requires no active wavefront modulators, diffractive optical elements, or complex multi-camera splitting architectures; instead, it relies entirely on the intrinsic effective birefringence and material dispersion of a single birefringent microscope slide. We experimentally demonstrate this simple multidimensional measurement modality using microscope slides made of calcite and lithium niobate ($LiNbO_3$), benchmarking its performance with immobilized fluorescent markers of well-defined polarization states and a pinhole array acting as a grid of point emitters back-illuminated at various central wavelengths to generate distinct spectral signatures. Furthermore, we showcase the practical utility of our method by performing a two-colour 3D localization of microparticles. By demonstrating robust multi-parameter retrieval from a single camera frame without any added hardware complexity, we outline a direct pathway toward super-resolution microscopy (SRM), showcasing how standard wide-field architectures can make multidimensional nanoscopy readily accessible to researchers without specialized optical expertise.

## RESULTS

By employing a uniaxial birefringent slide in a standard epi-fluorescence microscope, light emitted from a particle is split into two orthogonally polarized beams: the ordinary (o) and extraordinary (eo) beams (see Fig. 1(A,B)). While the o-beam obeys Snell's law, the eo-beam refracts at a walk-off angle $\rho$ governed by the crystal's birefringence and the angle $\alpha$ between its optical $c$-axis and the optical $z$-axis. Consequently, both beams undergo a lateral separation within the plane spanned by the $c$- and $z$-axes (here aligned with the $y$-axis of the imaging system). Depending on the walk-off angle and the slide thickness $D$, this separation yields two distinct, non-overlapping particle images – a double image – on the camera sensor with distance $\Delta Y$ in between. This is illustrated in Fig. 1(C) for a $D$ = 200 µm calcite slide ($\alpha = 45°$).

Crucially, while the lateral distance $\Delta Y$ obviously remains constant across the axial $z$-range, a bi-focal imaging results due to the different refractive indices ($n_o$, $n_{eo}$) and the walk-off of the eo-beam. The particle is thus imaged twice with different levels of defocus depending on its axial position relative to the focal planes $F_o$ and $F_{eo}$. Because the refractive index $n_{eo}$ of the eo-beam is direction dependent, its point spread function (PSF) exhibits pronounced astigmatic aberrations, elliptically distorting the corresponding particle image (see Fig. 1(C)). This results in two distinct axial foci, $F_{eo,x}$ and $F_{eo,y}$, along the $x$- and $y$-directions, respectively. The axial shifts ($\Delta z_x$, $\Delta z_y$) relative to $F_o$ are strictly governed by the thickness and birefringence of the slide, as experimentally confirmed by comparing the PSFs for three different birefringent plates (Fig. 1(D-F)). Increasing the thickness of a lithium niobate (LN) slide from $D$ = 273 µm to 507 µm proportionally scales both the axial shifts ($\Delta z_x$, $\Delta z_y$) and the lateral distance $\Delta Y$ (Fig. 1(D,E)). In contrast, a 200 µm calcite (CAL) slide – despite being thinner – results in a significantly larger axial and lateral displacements due to its much stronger birefringence (Fig. 1(F)). However, a very large axial shift can hinder the simultaneous detection of both particle images over a wide depth range, as a highly defocused image spreads over a large area making it difficult to distinguish from the background noise[41,42] (see e.g. the particle image of the o-beam at $F_{eo,x}$ depicted in Fig. 1(C)). Ultimately, while the lateral distance $\Delta Y$ can be adjusted by the system magnification and the physical properties of the birefringent microscope slide, the bifocality and the astigmatic axial shifts ($\Delta z_x$, $\Delta z_y$) are mainly governed by the birefringent substrate.

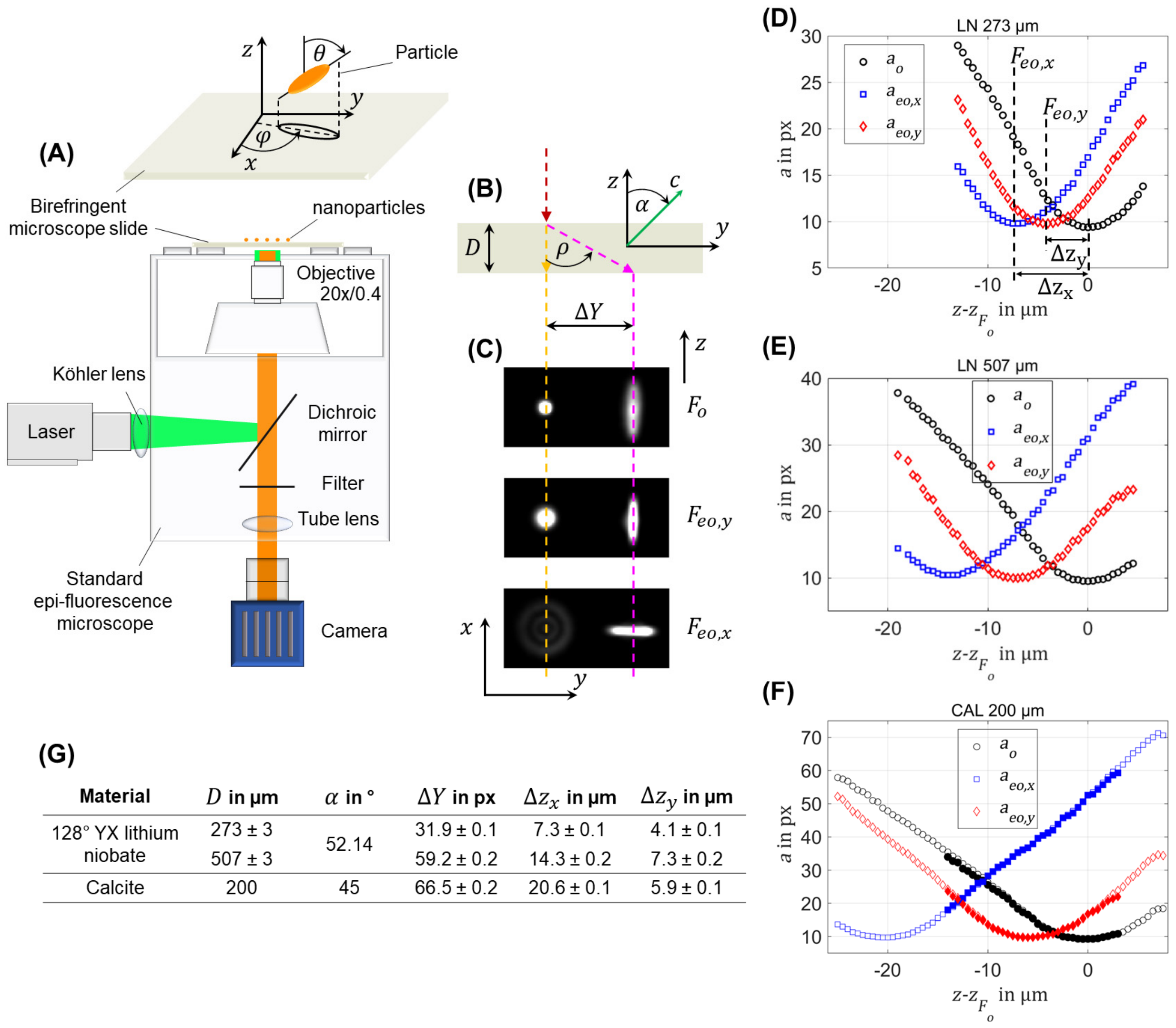

| Material | $D$ in µm | $\alpha$ in ° | $\Delta Y$ in px | $\Delta z_x$ in µm | $\Delta z_y$ in µm |
|---|---|---|---|---|---|
| 128° YX lithium niobate | 273 ± 3 | 52.14 | 31.9 ± 0.1 | 7.3 ± 0.1 | 4.1 ± 0.1 |
| | 507 ± 3 | | 59.2 ± 0.2 | 14.3 ± 0.2 | 7.3 ± 0.2 |
| Calcite | 200 | 45 | 66.5 ± 0.2 | 20.6 ± 0.1 | 5.9 ± 0.1 |

***Fig. 1 Optical configuration, operating principle, and characterization of DRM.*** *(A) Sketch of the optical setup consisting of a standard epi-fluorescence microscope in combination with a birefringent microscope slide, and a nanoparticle illustrated in 3D with specific orientation expressed by an azimuthal (φ) and polar (θ) angle. (B) Side view of the birefringent microscope slide of thickness D and its optical c-axis within the yz-plane at a tilt angle α. Light is split into an ordinary (o) and extraordinary (eo) beam, whereby the eo-beam is refracted under a walk-off angle ρ. (C) Both beams are imaged on a camera sensor causing a double image with a lateral distance ΔY in between. The particle image of the eo-beam shows astigmatic aberrations. The three examplary double images illustrate a nanoparticle of 450 nm in diameter imaged through a microscope slide made of calcite (CAL) with 200 µm thickness taken at three distinct positions. The three positions correspond to the focal plane of the o-beam ($F_o$) as well as the two foci of the eo-beam ($F_{eo,x}$, $F_{eo,y}$) caused by astigmatism. (D-F) Measured particle image diameters for the three different microscope slides used in this study made of calcite (CAL) and lithium niobate (LN) of different thicknesses. For all profiles, red-fluorescent nanoparticles of 450 nm where drop-cast above the birefringent microscope slide. In the case of CAL, the nanoparticles were imaged at low (filled markers, corresponding to the illumination conditions used also for LN) and high intensity (open spheres) to capture the profiles over a larger axial range. (G) Table listing the three different birefringent slides used with its respective thicknesses and tilt-angle α as well as the measured lateral distance ΔY and the axial displacements ($\Delta z_x$,$\Delta z_y$) of both foci ($F_{eo,x}$,$F_{eo,y}$) of the eo-beam with respect to the focal plane ($F_o$) of the o-beam. In the case of the CAL substrate, only the nominal thickness is known.*

As illustrated in Fig. 2(A), the double image yields a wealth of information featuring multidimensional measurement as follows.

First, the 3D position of particles can be determined by the lateral centroids ($X_o$, $Y_o$; $X_{eo}$,$Y_{eo}$) and the respective diameters of the ordinary and extraordinary particle images, which exhibit different levels of defocus depending on the axial $z$-position. As shown in Fig. 2(B), the difference of the particle image diameters squared in $x$-direction ($a_o^2 - a_{eo,x}^2$) uniquely relates to the $z$-position of the particle, independently of the particle's physical size[43]. Hence, this calibration function holds across both non-diffraction and diffraction-limited regimes, enabling robust 3D position estimation in polydisperse systems. Beyond the magnification and numerical aperture of the imaging system, the calibration function is highly dependent on the material properties of the birefringent microscope slide. Again, the thicker the microscope slide and the larger the birefringence, the steeper the calibration function (see Fig. 2(C)). Hence, the birefringent plate made of calcite exhibits the highest sensitivity for the $z$-position compared to the two microscope slides made of LN. For the latter, the slope of the calibration function relates to the thickness of the microscope slide.

Second, the imaging method is inherently polarization-sensitive, as the o- and the eo-beams constitute two orthogonal polarization channels. This unique property can be exploited to determine the azimuthal orientation ($\varphi$) of dipole emitters by calculating the linear dichroism (LD), defined as[21,44]:

$$LD = \frac{I_o - I_{eo}}{I_o + I_{eo}}$$

where $I_o$, $I_{eo}$ represent the intensities of the respective double images (see Fig 2(A)). To validate this orientation sensitivity, red-fluorescent particles were immobilized on a linear polarizer, simulating single dipole emitters with well-defined azimuthal orientations relative to the polarization axis of the o-beam. For these validation measurements, a 273 µm LN slide was positioned between the polarizer and the microscope objective. As illustrated in Fig. 2(D,E), the intensities of the double image change distinctively as a function of the polarizer angle, confirming that the azimuthal orientation can be directly inferred from the $LD$. However, due to the bi-focal imaging, the particle image intensities, and thus the resulting $LD$, depend on the particle's axial position. This dependency is demonstrated in Fig. 2(E) for two distinct $z$-positions coinciding with the intensity maxima of the o- and eo-beam (see Fig. 2(D)). Therefore, precise prior knowledge of the emitter's 3D position is a mandatory prerequisite for the accurate determination of its azimuthal orientation.

Third, the lateral distance $\Delta Y$ of the double image is inherently wavelength-dependent, enabling the direct determination of the spectral signature. As illustrated in Fig. 2(F), imaging a two-colour back-illuminated pinhole through a 273 µm LN slide yields three distinct spots in total. Under simultaneous excitation at central wavelengths of 402.0 nm and 631.7 nm, the particle images of the o-beam superimpose perfectly, whereas the particle images of the eo-beam separate due to a chromatic walk-off. Consequently, the lateral distance $\Delta Y$ can be exploited to decode the spectral signature of an emitter, e.g. to distinguish between different fluorophores. The wavelength-dependent lateral distances $\Delta Y$ for the three-different birefringent slides used in this study are depicted in Fig. 2(G). For this characterization, a pinhole array served as a grid of point emitters, whose spectral characteristic was controlled either by selecting one of the six LEDs in the SpectraX light source (filled markers in Fig. 2(G)) or by employing a tunable white light laser (unfilled markers). As expected, $\Delta Y$ monotonically decreases with increasing wavelength in all three cases. Notably, LN exhibits a significantly higher chromatic dispersion of birefringence than calcite, yielding a much stronger spectral sensitivity of the lateral distance $\Delta Y$. This is clearly highlighted in Fig. 2(H) by the normalized lateral distance $\Delta Y_n$ relative to the value at 402 nm.

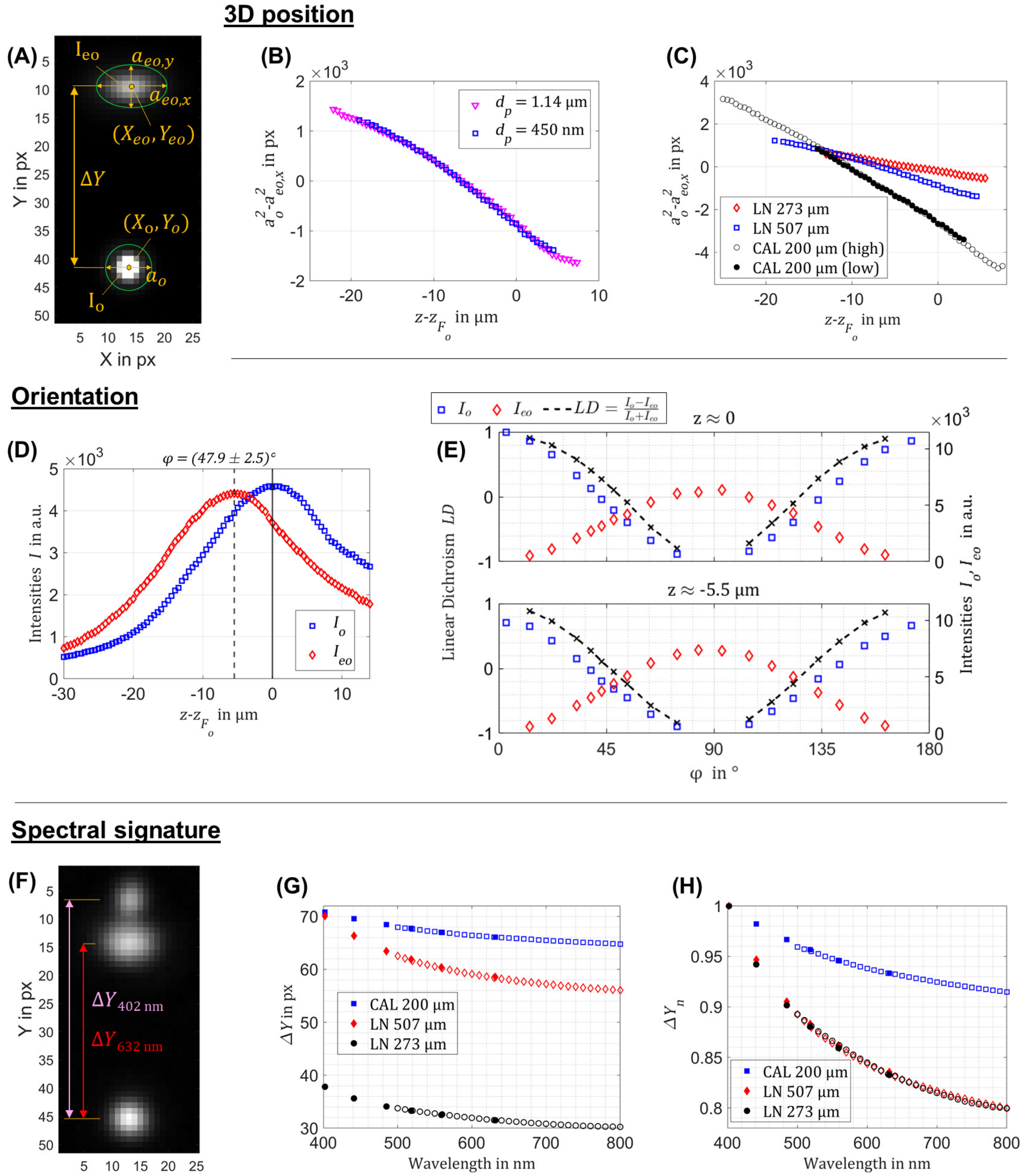

*Fig. 2* ***Calibration and characterization of DRM for 3D Position, Orientation, and spectral signature in DRM.*** *(A) Example of a double image of a nanoparticle with 450 nm in diameter, and indication of the measured quantities giving access to the measurement of 3D position, orientation and spectral signature of nanoprobes. (B) Calibration function of the axial position based on the squared difference of the particle image diameters of the o-beam ($a_o$) and the eo-beam in x-direction ($a_{eo,x}$). The results obtained for two red fluorescently labelled particles of different sizes confirm its uniqueness from the non- to the diffraction-limited case. (C) Calibration functions for the axial position for the three different birefringent slides used in this study under identical illumination conditions. The open black spheres represent a measurement with higher intensity, demonstrating also the intensity-independency of the calibration function. (D) Measured intensities ($I_o$, $I_{eo}$) for the double images of an emitter with linear polarization state measured at a specific orientation with respect to the orientation of the polarization of the o-beam and along the axial direction. (E) Measured intensities of the double images and linear dichroism (LD) for different orientations of the emitter at the two specific z-positions. (F) Image taken from a point emitter back-illuminated with a multispectral LED, simultaneously emitting light at a central wavelength of 402 nm and 632 nm. (G) Lateral distances $\Delta Y$ determined from a grid of point emitters back-illuminated either with a multispectral LED (filled markers) or white light laser with narrow spectral bandwidth (open markers), to cover a large wavelength range from dark blue to near infrared for all three types of birefringent slides used in this study. (H) Relative lateral distance $\Delta Y_n$ normalized with the lateral distance at 402 nm wavelength, illustrating the relative change with wavelength. Consistent with the underlying calculated material characteristics (see Fig. 4), LN exhibits a substantially stronger dispersion of birefringence compared to CAL, rendering it inherently more sensitive for resolving the spectral signatures of emitters.*

**Two-colour 3D localization measurement**

To demonstrate the practical utility of our new imaging method and evaluate the impact of spatio-spectral coupling, we designed a two-colour localization benchmark inside a microfluidic gap of height $H$ = 9.98 µm (established via monodisperse, plain $SiO_2$ spacer particles, see Fig. 3(A)). The setup utilized two distinct, monodisperse particle populations matched to a nominal diameter of ~ 2.44 µm: heavy, blue-fluorescent $SiO_2$ particles and light, red-fluorescent polystyrene (PS) particles suspended in an intermediate-density aqueous glycerol solution. Microparticles were chosen strategically: unlike nanoparticles, their sedimentation (for $SiO_2$) and buoyancy (for PS) dominate over the Brownian motion. This allows them to self-assemble into static reference planes at the bottom and top of the gap, while providing an excellent signal-to-noise ratio (SNR) for precise three-dimensional particle image characterization.

The chromatic walk-off of the birefringent slide generates double images (Fig. 3(C)) whose lateral separation $\Delta Y$ serves as a unique spectral signature, enabling unambiguous classification of both populations at any depth (Fig. 3(D)). However, the calibration curves reveal a distinct dispersion effect: while $\Delta Y$ remains nearly constant for the red PS particles, the blue $SiO_2$ particles exhibit a depth-dependent lateral distance $\Delta Y$. This behavior is a direct manifestation of spatio-spectral coupling. Specifically, the broad emission bandwidth of the blue fluorophore interacts with the strong material dispersion of the LN slide in this spectral window (see Fig. 4(A)), translating a longitudinal chromatic splitting along the optical axis into the observed depth-dependent lateral drift $\Delta Y$ during the $z$-scan. Because material dispersion is significantly lower in the longer wavelength regime, this coupling remains imperceptible within the standard deviation of $\Delta Y$ for the red-fluorescent PS particles. Crucially, as the $\Delta Y$ domains remain widely separated across all depths, particle species classification is preserved with absolute fidelity.

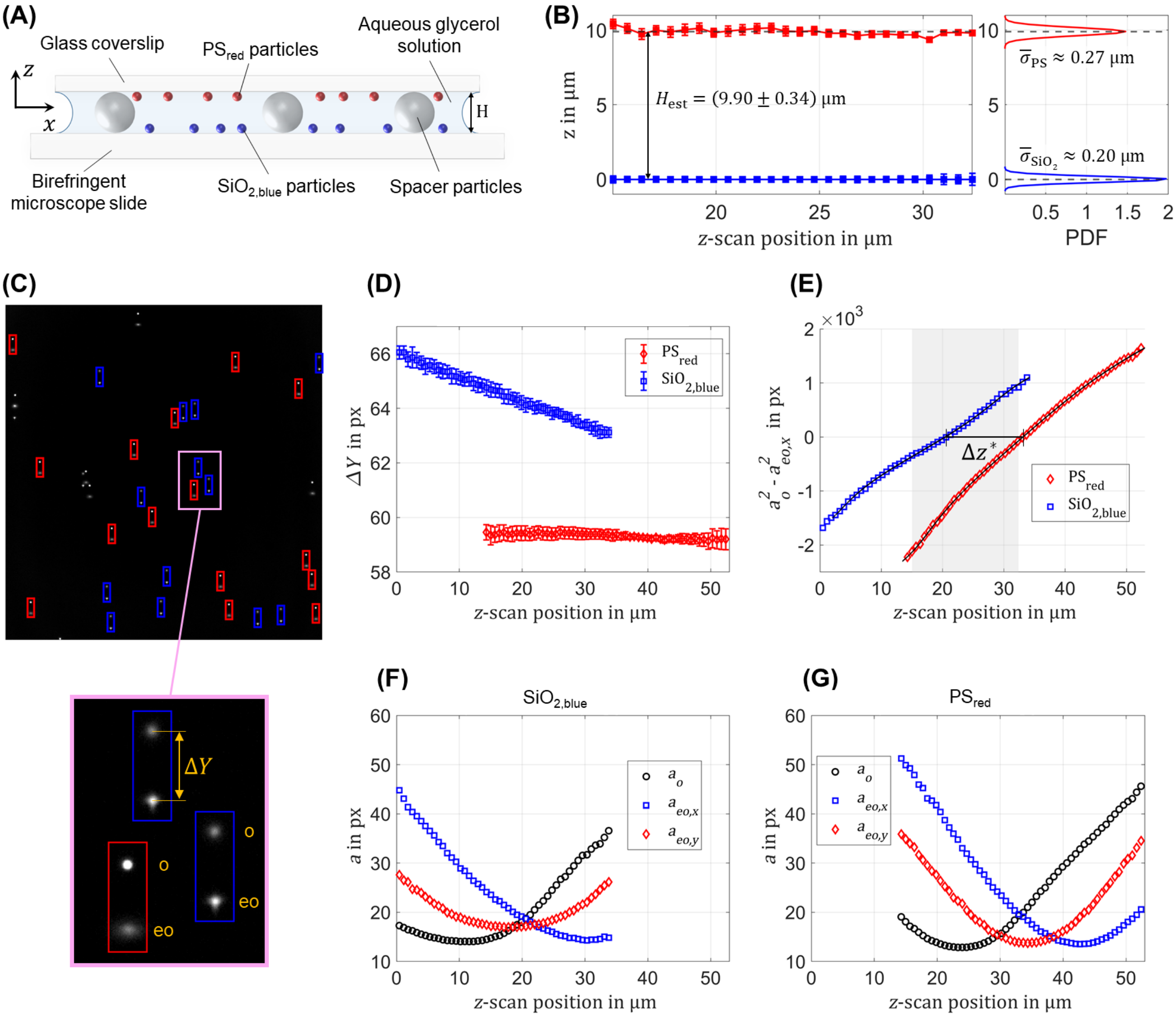

*Fig. 3* ***Two-colour localization proof-of-concept experiment in a microgap demonstrating spatio-spectral dispersion.*** *(A) Sketch of the two-colour localization experiment consisting of the birefringent microscope slide made of LN with 507 µm thickness and a glass coverslip located above (not to scale). The gap height H between both slides was set by spacer particles with a diameter of (9.98 ± 0.34) µm. Heavy, blue-fluorescent $SiO_2$ particles and light red-fluorescent PS particles were placed at the bottom and top of the microgap, respectively. (B) Measured locations for both particle species determined at z-scan positions within 15 µm to 32 µm. At each scan position, the location of the heavy, blue-fluorescent particles was set to zero as reference. The error bars indicate the standard deviation of the mean values at each position. On average, these standard deviations amount to 0.20 µm and 0.27 µm for the blue- and red-fluorescent particle species, respectively. The estimated gap height amounted to $H_{est}$= (9.90±0.34) µm. (C) Example image captured at z-scan position 21.97 µm. According to the zoom-in view, the blue- and red-fluorescent particle images show different characteristics and can be discriminated by the lateral distance ΔY in between the double image. (D) Determined lateral distance ΔY along the z-scan position for the blue-fluorescent $SiO_2$ and red-fluorescent PS particles. (E) Calibration functions for the axial position for both particle species. Both curves are separated from each other because of the different height of both particle species. The grey shaded area indicates the measurement position where the gap height H has been determined. (F+G) Measured particle image diameter profiles for both particles species, demonstrating pronounced spatio-spectral coupling within the blue emission regime.*

The experimental calibration functions, depicted in Fig. 3(E), as a function of the effective z-scan position (scaled by the fluid's refractive index to reflect true axial dimensions), show a distinct separation. This separation primarily reflects the microgap configuration: the heavy, sedimented blue-fluorescent $SiO_2$ particles are encountered early in the *z*-scan, whereas the buoyed red-fluorescent PS particles at the upper coverslip appear at larger scan positions. Due to chromatic dispersion, the fluorophores exhibit slightly different calibration curves for the *z*-position; thus, while the calibration function is independent of the particle size, individual curves must be acquired for each fluorescent marker. Nevertheless, establishing both calibration functions with a common $z$-scan enable a quantitative baseline for the two-colour localization. For instance, the axial distance $\Delta z^*$ between the zero-crossings of the two curves (representing identical o- and eo-image diameters in $x$-direction of the double image) yields an apparent distance of approximately 12.69 µm. Crucially, the longitudinal chromatic focal shift of $z_{\mathrm{cs}} \approx 5.23$ µm between both emission bands artificially increases the optically measured distance. This chromatic shift was independently quantified using a single-plane reference configuration (see SI, Note 1). By correcting this offset and accounting for the mean particle diameters of both particle species, the reconstructed physical gap height is $H_{\mathrm{est}} \approx 9.90$ µm, which is in excellent agreement with the nominal spacer particle diameter of (9.98 ± 0.34) µm. The reconstructed gap height remains remarkably constant, fluctuating tightly around the mean value of 9.90 µm, with an overall standard deviation of 0.34 µm derived from the positional variations of both particle species across the entire evaluation range ($z \approx 15$ µm to 32 µm, shaded grey in Fig. 3(E)). This demonstrates that our dual-colour axial localization remains highly precise and depth-invariant throughout the entire volume.

This position-independent accuracy is achieved by extracting the axial coordinate exclusively from the particle image diameters along the $x$-direction ($a_{ao,x}$). As shown in Fig. 3(G,H), these $x$-directed profiles retain well-defined, symmetric defocusing curves with distinct focal minima across both colour channels. This symmetry arises because they are significantly less affected by the directional dependency of the extraordinary refractive index. Conversely, the extraordinary particle image diameter along the walk-off direction ($a_{eo,y}$) also maintains a symmetric profile around the focus but exhibits a significantly broader curve and an increased overall diameter. While this effect is highly pronounced in the blue wavelength regime (Fig. 3(G)), the red channel (Fig. 3(H)) shows only a minor widening compared to $a_o$ due to the lower material dispersion at longer wavelengths. Physically, this broadening stems from the angular dependence of the extraordinary refractive index $n_{eo}$ within the objective's numerical aperture. Since the optical axis of the substrate lies in the $yz$-plane, the resulting wavefront aberrations affect both lateral axes but are dominantly manifested along the $y$-direction.

This results in additional focal blurring in the $y$-direction, while remaining negligible in the $x$-direction. For the blue-fluorescent particles, this effect is further amplified by the broad spectral bandwidth of the fluorophore, which couples with the directional refractive index gradients to generate the observed larger particle image diameter $a_{eo,y}$. Consequently, partitioning image feature extraction – utilizing the diameters in $x$-direction for axial localization while reserving the $y$-directed features for species classification – effectively reduces the impact of dispersion-induced aberrations on spatial reconstruction, mitigating potential depth-dependent localization errors.

## DISCUSSION

The successful sub-micrometer reconstruction of the microgap height underpins a crucial physical insight: the inherent anisotropic and dispersive effects of a birefringent substrate enable deterministic features that can be leveraged for multidimensional microscopy. This demonstrates that Double Refraction Microscopy (DRM) can access the full dimensional information of an emitter by resolving axial depth, spectral signature, and potentially dipole orientation. Unlike conventional multi-channel methods that externally split the emission light into separate detection paths, DRM operates as a passive, inline optical encoder without degrading the signal-to-noise-ratio through lossy optical splitters. By capturing all multidimensional features within a single, un-split detection path, it retains the full photon budget – the primary prerequisite for high localization precision – while completely eliminating hardware-induced channel misalignment. Furthermore, DRM avoids high hardware costs associated with active wavefront-shaping devices such as spatial light modulators.

However, to transition DRM towards true multidimensional super-resolution microscopy, several key challenges must be addressed. First, the selection of thin substrates with tailored birefringence is essential to achieve appropriate bifocality and sufficient image separation, especially when dealing with the limited photon budget of single molecules. Second, the aberrations expected to arise in both the ordinary and extraordinary beams will require systematic characterization and mitigation through precise substrate design. In the case of the extraordinary beam, these aberrations mainly stem from the direction-dependent refractive index, a unique physical aspect that inherently modulates the wavefront across the wide angular spectrum of high-numerical-aperture (NA) objectives. Here, the optimal choice of the material parameters, including birefringence, absolute refractive indices, substrate thickness and crystal orientation, is decisive. However, this challenge also presents a distinct opportunity: the direction-dependence could potentially translate into an enhanced sensitivity for measuring 3D dipole orientation – a hypothesis that remains to be investigated. Third, while utilizing a static, pre-optimized birefringent slide limits experimental flexibility compared to conventional methods, system-level strategies can be employed to mitigate this limitation. For instance, narrow bandpass filters can mitigate dispersion-induced distortions, particularly for the extraordinary beam, albeit at the expense of the photon budget[30]. Alternatively, inherently narrow-band emitters, such as quantum dots, can be utilized to circumvent these dispersive aberrations without compromising the photon budget[30,33]. Fourth, while the current image processing and evaluation successfully demonstrates the feasibility of DRM, they remain unoptimized. Leveraging state-of-the-art multi-parameter estimation algorithms could significantly push the method toward its fundamental limits. Specifically, implementing maximum-likelihood estimation (MLE) based on physically simulated PSF models[45,46], which are essential to account for the complex, non-Gaussian profiles induced by dipole emitters[47], or integrating deep-learning-assisted multi-emitter fitting routines would maximize the estimation precision and robustness of the multi-dimensional parameter extraction[48,49].

Ultimately, the core feature of DRM is its remarkable experimental simplicity. Operating as a passive, single-camera add-on that requires no intricate optical alignment, it provides a straightforward pathway to multidimensional super-resolution microscopy for researchers without specialized optical expertise or access to cost-prohibitive instrumentation. Due to this inherent versatility, the method seamlessly integrates into both fluorescence and dark-field microscopy modalities, unlocking a comprehensive multi-parameter space across diverse scientific domains. For instance, the spectral signature enables the monitoring of binding processes between differently labeled molecules in life sciences, while – in material sciences – it serves as an intrinsic metric to determine size, shape, and material of plasmonic nanoprobes. Therefore, the combination of experimental simplicity and multi-parameter capabilities establishes DRM as a versatile platform to advance research in nanoscience.

## METHODS AND MATERIALS

### Birefringent elements

In total, three distinct birefringent substrates were investigated in this work: one commercial calcite crystal plate and two custom lithium niobate ($LiNbO_3$) slides of different thicknesses (see Fig. 1(G)).

The calcite sample (P/N: CAL12020-45-AR800/400, NewLight Photonics, Canada) featured a 12.7× 12.7 mm² aperture and a nominal thickness of 200 μm. The crystal possesses an $\alpha$= 45° cut orientation and is equipped with a dual-band anti-reflection (AR) coating optimized for 400 nm and 800 nm. The remaining two samples were custom $LiNbO_3$ microscope slides with distinct thicknesses, fabricated by dicing commercial wafer material. These substrates consisted of surface acoustic wave (SAW) grade, double-side polished 127.86°-rotated Y-cut $LiNbO_3$ crystals. The wafers with 507 μm and 273 μm thicknesses were supplied by Siegert Wafer (Germany) and Hangzhou Freqcontrol Electronic Technology Ltd (China), respectively. The actual thicknesses of the diced microscope slides were mechanically characterized by taking ten independent measurements across different spatial positions using a high-precision dial indicator.

### Pinhole array fabrication

The pinhole array was fabricated using a commercial 5" × 5" × 0.09" soda-lime glass mask blank (MB Whitaker & Associates, Germany). The blank was pre-coated with a 100 ± 10 nm low-reflective chromium layer and a 530 nm thick film of AZ 1518 positive photoresist. Lithographic patterning was performed using a maskless aligner (MLA150, Heidelberg Instruments, Germany) with an exposure dose of 95 mJ/cm² at zero defocus. The pinholes had a nominal diameter of 2.5 μm and were arranged in a staggered array with a vertical repeating distance of 50 μm (along the $y$-direction) and a horizontal pitch of 25 μm (along the $x$-direction). Following exposure, the photoresist was developed in AZ 351B (Microchemicals GmbH, Germany) for approximately 40 s. The underlying chromium layer was then patterned via wet-chemical etching in a commercial chromium etchant for 2 min. After etching, the remaining photoresist was stripped wet-chemically using AZ 100 remover at 60°C. Finally, the mask was diced into the final substrate dimensions using a dicing saw equipped with a specialized Microkerf blade (model 2.187-12-45S21). To ensure a clean edge and prevent substrate damage, the total dicing depth of 2450 µm was reached progressively in three sequential cutting steps.

**Tracer and spacer particles**

All particles utilized in this study were purchased from microParticles GmbH (Berlin, Germany). Three different sizes of red-fluorescent polystyrene (PS) particles were employed, featuring nominal excitation and emission maxima of 530 nm and 607 nm, respectively. The mean particle diameters and standard deviations were (450 ± 9) nm (lot no. PS-FluoRed-Fi144-2), (1.14 ± 0.030) µm (lot no. PS-FluoRed-Fi329), and (2.47 ± 0.040) µm (lot no. PS-FluoRed-Fi235).

Additionally, blue-fluorescent silica $SiO_2$ particles with a nominal diameter of (2.41 ± 0.33) µm (lot no. $SiO_2$-FluoBlue-L5040) were used, featuring nominal excitation and emission maxima of 350 nm and 450 nm, respectively. To precisely control the microgap height, non-fluorescent, plain $SiO_2$ particles with a nominal diameter of (9.98 ± 0.34) µm (lot no. $SiO_2$-F-SC2222-2) were used as spacers.

To characterize the specific spectral ranges of the fluorophores susceptible to dispersion effects in the birefringent material, the spectral signatures of the fluorescent particles were characterized using a spectrometer (SensLine, Avantes B.V, The Netherlands). Excitation was provided for the blue-fluorescent and red-fluorescent particles by a UV LED (SOLIS-365, Thorlabs Inc., USA) and a 532 nm laser (FP-D-532-10-C-F, BLAU Optoelektronik, Germany), respectively. Long pass filters with cut-on wavelength of 425 nm (#84-742, Edmund Optics BV, The Netherlands) and 550 nm (FELH0550, Thorlabs Inc.) were used for the blue- and red-fluorophore, respectively. The resulting emission spectra are plotted alongside the wavelength-dependent effective birefringence ($\Delta n_{\text{eff}}$) of $LiNbO_3$ in Fig. 4(A). For simplicity, $\Delta n_{\text{eff}}$ was calculated assuming plane wave propagation parallel to the system's optical axis (normal incidence) based on the known orientation of the crystal's $c$-axis.

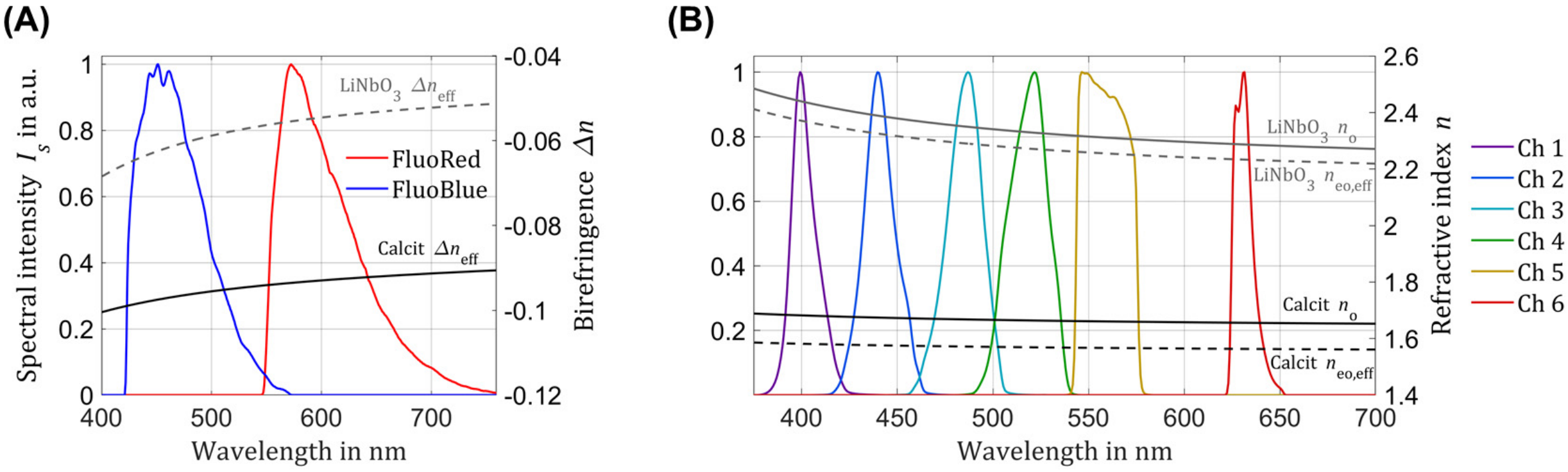

***Fig. 4 Spectral characterization of the fluorophores and LED illumination alongside calculated material dispersion**. (A) Emission spectra of the blue- and red-fluorescent microparticles displayed alongside the wavelength-dependent effective birefringence ($\Delta n_{eff}$) of LiNbO3 and calcite. The center wavelengths, determined via centroid calculation, are 486.5 nm for the blue and 604.5 nm for the red particles, respectively. (B) Spectral profiles of channels 1–6 (Ch1–Ch6) of the SpectraX LED light source, overlaid with the calculated effective refractive indices of LiNbO3 and calcite. The corresponding centroid wavelengths are characterized as 402.0 nm (Ch1), 441.1 nm (Ch2), 484.9 nm (Ch3), 518.2 nm (Ch4), 558.9 nm (Ch5), and 631.7 nm (Ch6). The illustration of the refractive indices and birefringence relative to these utilized wavelengths directly explains the experimental results: while calcite provides a substantially higher birefringence, LiNbO3 exhibits a significantly more pronounced dispersion in both its refractive indices and its birefringence across the investigated spectral range. In both panels, 'effective' denotes material properties calculated for paraxial rays propagating parallel to the system's optical axis.*

## Calculation of refractive indices and birefringence

The wavelength-dependent ordinary and extraordinary refractive indices, as well as the resulting effective birefringence $\Delta n_{\text{eff}}$ for calcite and $LiNbO_3$, were calculated using their respective Sellmeier equations (see Supplementary Note 2 for the explicit equations and coefficients). All calculations were performed assuming paraxial plane wave propagation parallel to the system's optical axis, utilizing the known angular cutting orientation of the crystal's $c$-axis to determine the effective optical response across the investigated spectral range.

## Optical configurations and measurement procedures

The core optical setup (see Fig. 1(A)) consisted of a standard epi-fluorescence microscope (Axio Observer 7, Zeiss GmbH, Germany) equipped with a long distance Plan-Neofluar objective (M20x, NA = 0.4, Zeiss GmbH, Germany) and a scientific CMOS camera (imager sCMOS, LaVision GmbH, Germany). The three birefringent substrates used in this study were placed in front of the microscope objective. Their actual position, the light sources and optical filter set of the microscope as well as the measurement procedure were adapted to the four experimental configurations as follows.

*3D Position*

The experimental setup for the calibration measurements for the 3D position is sketched in Fig. 1(A). The birefringent substrates served as the microscope slide at which red-fluorescent polystyrene particles of 450 nm in diameter were drop-cast. A modulatable OPS-Laser (Tarm Laser technologies tlt GmbH & Co.KG, Germany) was used to illuminate the particles at a central wavelength of 532 nm under identical conditions applying an illumination duration of 1 ms. In the case of the high intensity measurements with Cal 200 µm (see Fig. 1(F) and 2(C), results illustrated with open markers), the laser was used at maximum power applying an illumination duration of 2 ms. To demonstrate the size-independent uniqueness of the calibration function (see Fig. 1(B)), additional calibration measurements were done with red-fluorescent polystyrene particles of 1.14 µm in diameter drop-cast on top of the microscope slide made of LN 507µm. In all cases, the microscope was equipped with a filter cube (LaVision GmbH, Germany) consisting of a bandpass filter (532±5) nm, a long pass dichroic mirror and a long pass filter with a cut-on wavelength of 540 nm and 542 nm, respectively. For all calibration measurements, the microscope objective was traversed precisely with equidistant step size of 0.5 µm in between each axial position. At each $z$-position, one image was taken.

*Orientation*

To demonstrate the orientation sensitivity, red-fluorescent polystyrene particles of 2.47 µm in diameter were drop-cast on a linear polarizer of 3.3 mm thickness (LPVISE100-A, Thorlabs Inc., USA), which was hold in place by a home-made 3D printed specimen holder. The birefringent slide made of LN 273 µm was placed below, enabling a free rotation of the polarizer, see Fig. 5(A). To adjust its orientation with respect to the polarization axis of the ordinary beam, the polarizer with particles was aligned manually by incrementing the orientation angle of about 5 degrees. The actual step size in between was estimated using the rotation of the static particle image pattern of the drop-cast particles, see Fig. 5(B,C). However, the relatively unknown initial orientation of the polarizer with respect to polarization axis of the o-beam causes a comparably high uncertainty estimated to be about 2.5°. For illumination, a high-power LED (SOLIS-525D, Thorlabs Inc., USA) was used. The microscope was equipped with the same filter set as described above for the case of the measurements for the 3D position. To determine the depth- and orientation dependent intensities of both images ($I_o$, $I_{eo}$) as well as the linear dichroism ($LD$) derived thereof (see Fig. 2(D,E)), images were taken at different axial positions by precisely translating the microscope objective with a minimum step size of 0.5 µm.

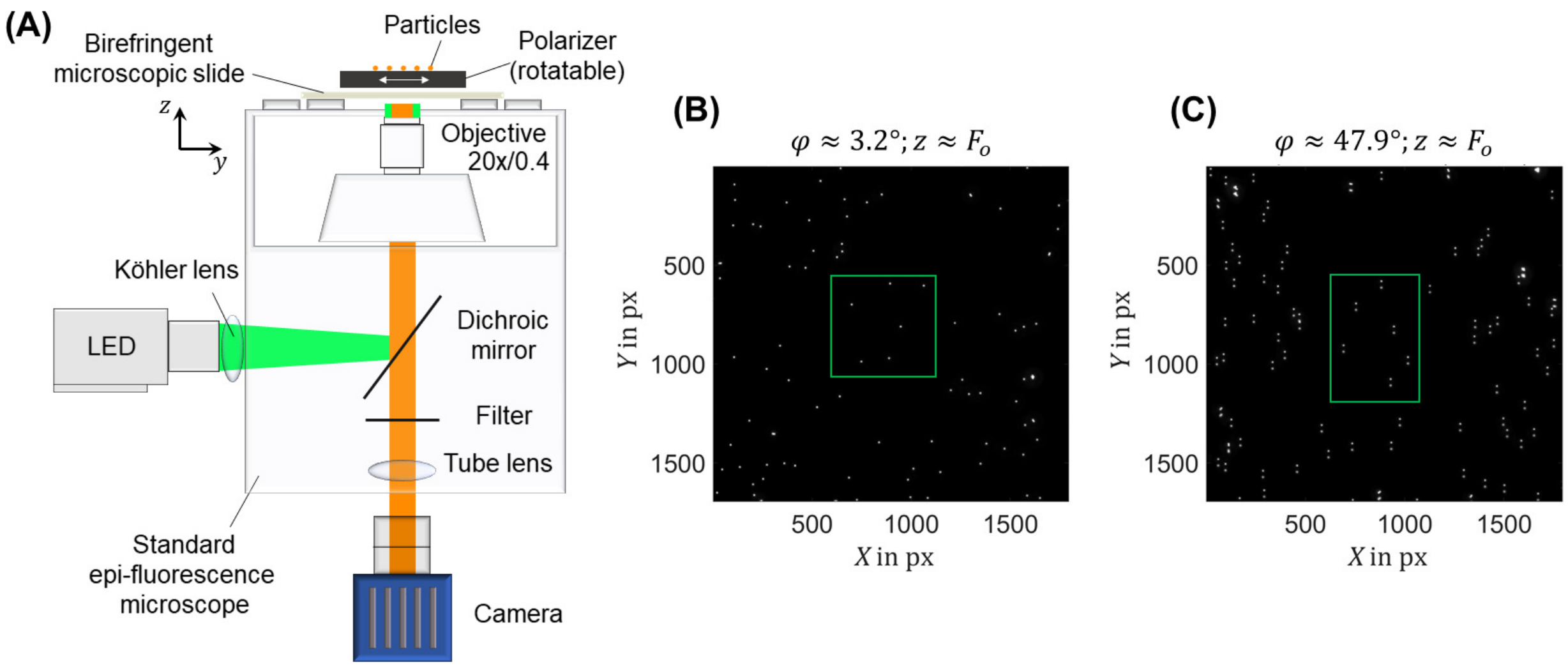

***Fig. 5 Evaluation setup and example images of orientation***. *(A) Schematic of the experimental setup used to demonstrate the measurement of the azimuthal orientation $\varphi$ of emitters by exploiting the inherent polarization of the o- and eo-beam. Red-fluorescent particles were drop-cast onto a linear polarizer to simulate emitters with linear polarization state. (B,C) Image captures of the drop-cast particles near the focal plane of the ordinary beam for two different polarizer orientations relative to the polarization axis of the ordinary beam. The particle pattern enclosed in the green boxes was used to determine the manually prescribed orientation. While only the particle image of the ordinary beam is visible in (B), the double images are discernible in (C). For enhanced visibility, the contrast was adjusted.*

*Spectral signature*

The experimental setup for qualifying the new method for the measurement of the spectral signature of emitters is depicted in Fig. 6(A). The pinhole array was placed face-down on top of the birefringent microscope slide, bringing the microfabricated pinholes into direct contact with the birefringent substrate. In the case of Cal 200 µm, the birefringent substrate was placed underneath the specimen holder. The pinhole array was back-illuminated from above utilizing a multispectral LED (SpectraX, Lumencor, USA) with six different central wavelengths. Their spectral characteristics are depicted in Fig. 4(B), measured by utilizing a spectrometer (SensLine, Avantes B.V, The Netherlands). In addition, a supercontinuum white light laser (SuperK EXTREME, NKT Photonics, Denmark) was used to extend the wavelength range. This laser was further equipped with a variable optical filter (SuperK Varia, NKT Photonics, Denmark) enabling to reliably select a narrow spectral bandwidth (10 nm) at the central wavelength between 500 nm and 800 nm. The step size in between was 10 nm.

In both with laser and LED illumination, a diffusor plate (DG20-1500, Thorlabs Inc., USA) was placed above the pinhole array, to suppress unwanted interference fringes and diffraction artifacts. To account for the dispersion-induced axial focal shift caused by the birefringent substrate, measurements were performed by translating the microscope objective with a step size of 0.5 µm. The lateral distance $\Delta Y$ was subsequently determined at the specific $z$-position where both particle image diameters ($a_o$, $a_{eo,x}$) are comparable. The intensity and illumination time of the light sources were tailored to each wavelength to achieve high-contrast double images.

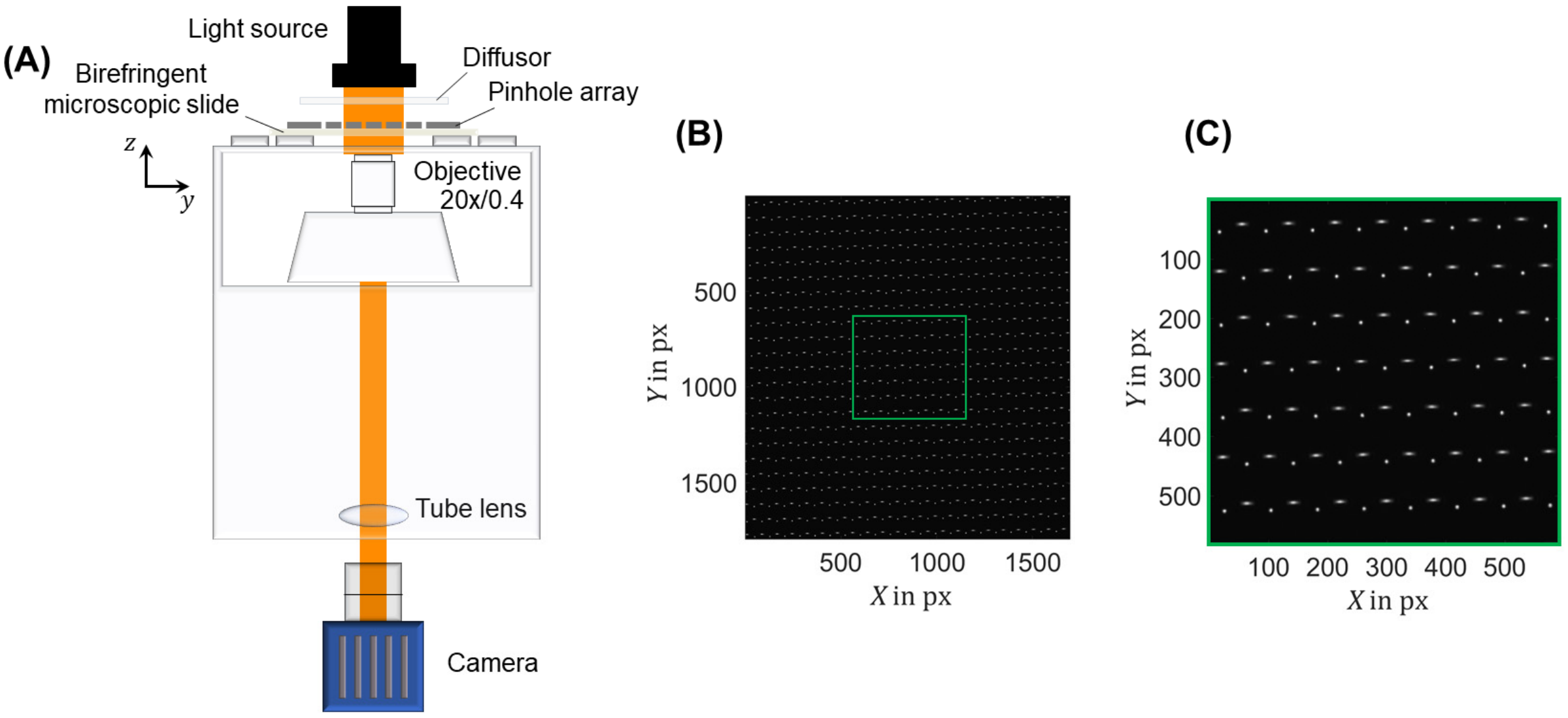

***Fig. 6 Evaluation and example images of spectral signature.*** *(A) Schematic of the experimental setup utilized to demonstrate the measurement of the emitter spectral signatures by exploiting the inherent dispersion of birefringence. (B) Wide-field image capture of the pinhole array, back-illuminated by a SpectraX LED light source (Ch6, centre wavelength 631.7 nm) and imaged through the 200 µm thick calcite slide (Cal 200µm). (C) Zoom in view of the double images within the highlighted green box in (B). The particle images corresponding to the o- and eo-beam are clearly distinguishable due the distinct astigmatism of the extraordinary beam.*

### *Two-colour 3D localization*

The experimental setup was identical to the optical configuration illustrated in Fig. 1(A), with the exception that the birefringent microscope slide with dried particles was replaced by the microgap assembly shown in Fig. 3(A). This gap was formed by a birefringent microscope slide made of LN 507 µm and an overlying glass coverslip, held at a well-defined distance $H$ by plain $SiO_2$ spacer particles and capillary forces. The gap was filled with a 40 wt.% aqueous glycerol solution, yielding a density of approximately 1100 kg/m³. To this working fluid, two distinct particle species were added at a low volume fraction to ensure individual and immobilized particles at the bottom and top surfaces of the microgap: blue-fluorescent $SiO_2$ at the bottom and red-fluorescent polystyrene particles at the top surface. To simplify the optical setup, illumination was provided by a single high-power UV-LED (SOLIS-365C, Thorlabs Inc., USA) at a central wavelength of 365 nm to simultaneously excite both fluorophores. Although the red-fluorescent particles are nominally specified for green excitation, their secondary absorption band in the UV spectrum yielded a sufficient photon budget for high-contrast imaging without requiring a secondary light source. To discriminate between excitation and emission wavelengths, the microscope was equipped with a filter cube consisting of a dichroic mirror (#86-330, Edmund Optics BV, The Netherlands) and long pass filter (#84-742, Edmund Optics BV, The Netherlands) with cut-on wavelengths of 409 nm and 425 nm, respectively. To scan the microgap, the microscope objective was translated axially with a step size of 0.5 µm. At each axial position, an image was recorded with an exposure time of 75 ms to compensate for the relatively low emission intensity, especially of the blue fluorophore.

**Image processing and data evaluation**

*Image preprocessing*

To eliminate peripheral vignetting artifacts caused by the camera sensor exceeding the optical field of view, the raw images were cropped to a central region of interest (ROI). Following a background subtraction step, a Gaussian smoothing filter was applied. For this, a 5×5 pixel kernel with a standard deviation of $\sigma$ = 2 px was utilized. In the case of the pinhole array measurements, the filter was adjusted to a larger smoothing width of $\sigma$ = 5 px to suppress the more pronounced diffraction and interference artifacts inherent to this optical configuration.

*Particle detection and feature extraction*

Particle segmentation was initially performed using a global intensity threshold, which was individually optimized for each experimental configuration to maximize the detection yield. To robustly identify valid particle images, the detected regions were filtered using strict geometric criteria. These criteria included a minimum structural area for each particle image (treating the two components of the double image independently at this stage) as well as maximum allowable particle image diameters along both the $x$- and $y$-directions. The segmented particle images were then analyzed to determine their precise in-plane positions and spatial dimensions. While the lateral in-plane position was extracted via a Gaussian-weighted centroid estimation, the particle image diameters along the $x$- and $y$-directions were determined using a one-dimensional Gaussian interpolation along each axis. Afterward, individual particle images were paired into their respective double images by searching for counterparts along the y-direction within a predefined distance range, accounting for the specific birefringent microscope slide and the emission wavelength of the fluorophore. For the spectral signature validation experiments using the back-illuminated pinhole array, the double images were cropped and an autocorrelation was applied to precisely determine the wavelength-dependent lateral distance $\Delta Y$.

*Data validation*

To validate the detected particle images and exclude false positives, a multi-step post-processing routine was implemented. First, any bright agglomerates bypassing the geometric filters were eliminated using an intensity threshold derived from the single-particle statistics. Second, strict geometric constraints were applied to the detected particle image shapes; while the ordinary beam image was required to be symmetric and circular, the particle image of the extraordinary beam was evaluated using a standard validation routine established for astigmatic particle tracking velocimetry (APTV) following the protocol described in Cierpka et al.[50] When mapping the particle dimensions in the $(a_{eo,x}, a_{eo,y})$-parameter space, valid single particles are constrained to a unique, continuous calibration curve that parameterizes the axial $z$-position. For each candidate, the minimum Euclidean distance between its measured dimensions $(a_{eo,x}, a_{eo,y})$ and this reference was computed. Particles deviating by more than 5 to 7 px from this reference curve, depending on the specific experimental run, were eliminated from further analysis.

While image acquisition was done with Davis 10.2.1 (LaVision GmbH, Germany), all subsequent steps from image preprocessing to data validation and editing were performed using custom scripts in MATLAB (R2024a, The MathWorks Inc., USA).

ACKNOWLEDGMENTS

The authors are grateful to David Schreier and the support by the Center of Micro- and Nanotechnologies (ZMN) of TU Ilmenau for the preparation of the pinhole array and birefringent microscope slides.

AUTHOR CONTRIBUTIONS

Conceptualization, Methodology, Investigation, Formal Analysis, Visualization, Writing – Original Draft, J.K.; Resources, C.C.; Writing – Review & Editing, J.K. and C.C.

COMPETING INTERESTS

The authors declare the following competing financial interests: Patent applications covering the DRM technology described in this work have been filed by the Technische Universität Ilmenau (DE102024119241A1; application no. DE102026117486.8) and by J.K. (PCT publication no. WO 2026/008167).

DATA AVAILABILITY

The datasets generated and/or analyzed during this study are available from the corresponding author on reasonable request.

CODE AVAILABILITY

The core image processing and evaluation were performed using a proprietary Matlab toolbox currently under active development at our institution. Additionally, custom Matlab scripts were written specifically for data visualization and the presentation of the figures. Both the visualization scripts and the underlying framework are available from the corresponding author upon reasonable request.

# Supplementary Information

# Multidimensional Double Refraction Microscopy

Jörg König[1,2*] and Christian Cierpka[1,2]

[1]Institute of Micro- and Nanotechnologies, TU Ilmenau, Germany

[2]Institute of Thermodynamics and Fluid Mechanics, TU Ilmenau, Germany

*e-mail: Joerg.Koenig@tu-ilmenau.de

## Note 1. Determination of the longitudinal chromatic shift

To determine the chromatic shift induced by the birefringent lithium niobate (LN) microscope slide (thickness: 507 µm), blue-fluorescent $SiO_2$ and red-fluorescent polystyrene (PS) particles with a mean diameter of 2.44 µm were drop-cast onto the substrate. Their spectral characteristics are displayed in Fig. 4(A). The experimental configuration was identical to the setup depicted in Fig. 1(A), except for the light source; here, a high-power UV-LED (SOLIS-365C, Thorlabs Inc., USA) operating at a central wavelength of 365 nm was utilized to simultaneously excite both fluorophores. To isolate the emission signals, the microscope was equipped with a filter cube comprising a dichroic mirror (#86-330, Edmund Optics BV, The Netherlands) and a long-pass filter (#84-742, Edmund Optics BV, The Netherlands) with cut-on wavelengths of 409 nm and 425 nm, respectively. To record the depth-dependent particle image profiles, the microscope objective was translated axially in steps of 0.5 µm. At each axial position, an image capture was taken, and individual particle images were analyzed following the evaluation routine detailed in the Methods section. The characteristic lateral distance $\Delta Y$, inherent to the double-refracting substrate, allowed for a clear discrimination between the two fluorophores (see Fig. S1(A)), confirming the findings from the two-colour localization experiment in the Results section. The calibration curves for the particle image diameters were established (see Fig. S1(B)). From these, the material-induced chromatic shift, $z_{cs}$, within the LN slide was quantified by fitting a model symmetric to the focus to the ordinary beam, as this component remains entirely unaffected by the spatio-spectral coupling anomalies described in the main text.

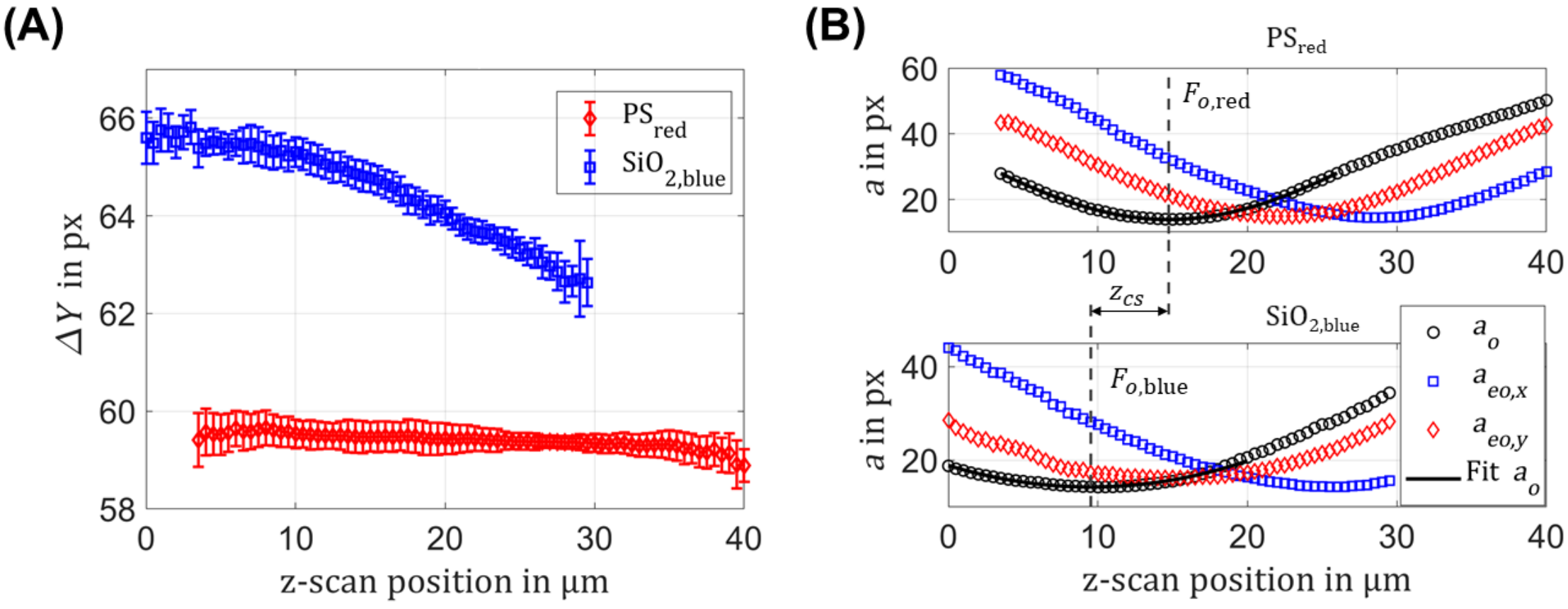

*Fig. S1 (A) Determined lateral separation distance $\Delta Y$ along the z-scan position for the blue-fluorescent $SiO_2$ and red-fluorescent PS particles. (B) Measured particle image diameters for both particle species (top: red-fluorescent PS, bottom: blue-fluorescent $SiO_2$). Indicated are the focal planes for the ordinary beam, from which the material-induced chromatic shift $z_{cs}$ was determined.*

**Note 2. Analytical calculation of refractive indices and dispersion**

To account for the wavelength-dependent optical response of the birefringent crystal plates used in this study, the ordinary ($n_o$) and extraordinary ($n_{eo}$) refractive indices were analytically calculated using established Sellmeier formulations. In all equations, the vacuum wavelength $\lambda$ is expressed in micrometers µm.

1. Dispersion coefficients for principal refractive indices

For calcite, the principal refractive indices were computed using the modified Sellmeier equation proposed by Ghosh [1]:

$$n^2(\lambda) = 1 + A + \frac{B\,\lambda^2}{\lambda^2 - C} + \frac{D\,\lambda^2}{\lambda^2 - E}$$

The specific numerical parameters utilized in the MATLAB routine are compiled in Table S1.

*Table S1 Sellmeier parameter for calcite according to Ghosh [1].*

| Parameter | Ordinary index ($n_o$) | Extraordinary index ($n_{eo}$) |
|---|---|---|
| ***A*** | 0.73358749 | 0.35859695 |
| ***B*** | 0.96464345 | 0.82427830 |
| ***C*** | $1.94325203 \times 10^{-2}$ | $1.06689543 \times 10^{-2}$ |
| ***D*** | 1.82831454 | 0.14429128 |
| ***E*** | 120.0 | 120.0 |

For congruent lithium niobate $LiNbO_3$, the principal refractive indices were calculated using the standard Sellmeier equation published by Zelmon et al. [2]:

$$n^2(\lambda) = 1 + \frac{A\,\lambda^2}{\lambda^2 - B} + \frac{C\,\lambda^2}{\lambda^2 - D} + \frac{E\,\lambda^2}{\lambda^2 - F}$$

The corresponding coefficients optimized for congruent material composition at room temperature are listed in Table S2.

*Table S2 Sellmeier parameters for congruent $LiNbO_3$ according to Zelmon et al. [2].*

| Parameter | Ordinary index ($n_o$) | Extraordinary index ($n_{eo}$) |
|---|---|---|
| ***A*** | 2.6734 | 2.9804 |
| ***B*** | 0.01764 | 0.02047 |
| ***C*** | 1.229 | 0.5981 |
| ***D*** | 0.05914 | 0.0666 |
| ***E*** | 12.614 | 8.9543 |
| ***F*** | 474.6 | 416.08 |

2. Derivation of effective extraordinary index and birefringence

Since the optical response depends on the propagation direction relative to the crystal's optic axis ($c$-axis), an effective extraordinary refractive index $n_{eo,\mathrm{eff}}$ must be derived. Assuming paraxial plane wave propagation parallel to the system's optical axis (normal incidence), the index ellipsoid equation simplifies to:

$$n_{eo,\mathrm{eff}}(\lambda,\alpha) = \frac{1}{\sqrt{\frac{\cos^2\alpha}{n_o^2(\lambda)} + \frac{\sin^2\alpha}{n_{eo}^2(\lambda)}}}$$

where $\alpha$ denotes the cutting angle between the crystal's $c$-axis and the system's optical axis. The specific cut angles used for the experimental configurations are:

- Calcite: $\alpha = 45°$
- Lithium Niobate: $\alpha = 52.14°$ (corresponding to a commercial $127.86°$ $Y$-cut SAW substrate)

Consequently, the wavelength-dependent effective birefringence $\Delta n_{\mathrm{eff}}(\lambda)$ is determined via:

$$\Delta \mathrm{n}_{\mathrm{eff}}(\lambda) = |n_{eo,\mathrm{eff}}(\lambda,\alpha) - n_o(\lambda)|$$